\documentstyle[twoside,fleqn,espcrc2,epsf]{article}

\newcommand{\cond}{\langle\bar\chi\chi\rangle}
\newcommand{\AmS}{{\protect\the\textfont2
  A\kern-.1667em\lower.5ex\hbox{M}\kern-.125emS}}

\title{Chiral Symmetry Breaking in the 3-d Thirring model for small $N_f$.}

\author{L. Del Debbio 
	\address{Centre de Physique Th\'eorique, 
        CNRS Luminy, F-13288 Marseille Cedex 9, France}%
	\thanks{This
        is work is supported by CNRS and the Angelo Della Riccia
        foundation.} (for the UKQCD Collaboration) }

\begin{document}

\begin{abstract}
We study the dynamical breaking of chiral symmetry in the 3-d Thirring
model for a small number of fermion species. The critical point is
identified by fitting lattice data to an equation of state. The
spectrum of the theory is studied to confirm the phase structure of
the model. 
\end{abstract}

\maketitle

\section{Introduction}

The Thirring model is a three dimensional QFT with $N_f$ fermion
species and a current--current four fermion interaction. As such, it
is obviously non--renormalisable in the usual perturbative
expansion. It has been pointed out several years ago~\cite{xxx} that a
continuum limit can be defined for some theories in a
non--perturbative (NP) approach. For the Gross--Neveu model, a
four--fermi theory with a (pseudo--)scalar interaction, the existence
of such an UV fixed point has been shown within the $1/N_f$ expansion
and confirmed by numerical simulations. The critical point separates a
strong coupling phase with chiral symmetry breaking from a symmetric
weak--coupling one.

For the Thirring model, solid conclusions can not be drawn from
analytic results as different NP techniques yield different results.
We aim to obtain a complete description of the phase structure of this
model by lattice simulation, using staggered fermions and a HMC
algorithm. The motivations, the lattice formulation of the theory and
the technical details of our analysis were presented in great details
in earlier works~\cite{ldd_and_sjh}, to which we refer the interested
reader. This contribution summarizes our most recent results for
$N_f=4,6$.

The order parameter for detecting chiral symmetry breaking is the
chiral condensate $\langle \bar\chi \chi \rangle$. Its behaviour
around the critical point is described by fitting to an equation of
state (EOS)~\cite{zinn93}. Finite--size effects are included in the
EOS and the critical exponents are obtained for $N_f=4$. The results
are compared with those for $N_f=2$. Once again, lattice data show a
clear signal of chiral symmetry breaking and fit rather well the RG
predictions.

Moreover the chiral symmetry breaking transition should manifest
itself in the behaviour of the susceptibilities and in the spectrum of
the theory. This issue was already addressed for $N_f=2$
in~\cite{ldd_and_sjh}. Preliminary results for $N_f=4$ are reported
here. The general pattern, which had already emerged in our previous
works, is confirmed by the new results. We briefly outline the next
developments in our program.

\section{Fitting to the Equation of State}

The EOS is a functional relation between the order parameter, the
coupling, $g$, and the bare mass, $m$, in the regularised theory,
which can be derived form the RG equations in the vicinity of a fixed
point (see e.g.~\cite{zinn93}). It therefore provides a relation to
which lattice data can be fitted. The results of this approach for
$N_f=2$ and fixed lattice size were presented
in~\cite{ldd_and_sjh}. In order to take into account finite size
effects, it is sufficient to consider the inverse linear size of the
lattice, $1/L$ as an additional scaling field of dimension 1. The
exact functional form of the EOS is unknown, but it can be expanded in
Taylor series, yielding a fitting function for the lattice data with
only a few unknown parameters:
\begin{equation}
	m = B \cond^\delta + A (t + C L^{-1/\nu}) \cond^{\delta-1/\beta} 
\end{equation}
where $t=1/g^2 - 1/g_c^2$. \\
For $N_f=4$, we fit to the EOS data from lattice whose sizes range
from $8^4$ to $16^4$. The results are summarized in Tab.~\ref{tab:fit}.

\begin{table*}[htb]
\caption{Results from fits including finite size scaling.}
\label{tab:fit}
\renewcommand{\thefootnote}{\thempfootnote}
\begin{tabular*}{\textwidth}{@{}l@{\extracolsep{\fill}}rrrrrr}
\hline
        & Parameter & Fit & & Parameter & Fit \\
\hline
$N_f=2$ & $1/g^2_c$ & 1.92(2) & $N_f=4$ & $1/g^2_c$ & 0.70(2) \\ &
        $\delta$ & 2.75(9) & & $\delta$ & 3.41(31) \\ &
        $\beta$ & 0.57(2) & & $\beta$ & 0.41(10) \\ &
        $\eta$ & 0.60(2) & & $\eta$ & 0.34(20) \\ 
        & $\nu$ & 0.71(4) & & $\nu$ & 0.61(10) \\   
        & $A$ & 0.334(7) & & $A$ & 0.70(5) \\ & $B$ & 2.7(3) & &
        $B$ & 5.8(2.1) \\ & $C$ & 2.1(7) & & $C$ & 2.9(2.2) \\ &
        $\chi^2$/d.o.f & 1.76 & & $\chi^2$/d.o.f & 0.80 \\
\hline
\end{tabular*}
\end{table*}
It is interesting to stress that the results from the fit do take into
account the effects due to the finite volume of the lattice and should
therefore be interpreted as values in the thermodymamical limit. The
values for the critical coupling and the critical exponents are stable
with respect to the previous results obtained for $N_f=4$ at fixed
lattice size~\cite{ldd_and_sjh}. Such a stability in our fits can be
seen as a sign that the lattice data really follow the behaviour
expected in the vicinity of an UV fixed point. The critical exponents
differ significantly between $N_f=2$ and $N_f=4$, confirming a result
that had already emerged from the fixed size fits. We should be able
to reduce the errors on the fitted values as statistics increase
over the next months. It is worthwhile to stress once again that, in
order to identify unambiguously a fixed point, we should study the
renormalised trajectories in coupling space, for which we have to
measure 3- and 4-point correlation functions.

\section{Susceptibilities and spectrum}

Independent information on the phase structure of the theory can be
obtained from the susceptibilities and the spectrum of the
theory. Once again precise definitions and the results for the $N_f=2$
case are given in previous works and we want to concentrate on the new
results for $N_f=4,6$. We briefly recall the definitions and the
expected behaviour for some quantities of interest.

The transverse and longitudinal susceptibilities are defined
respectively as:
\begin{eqnarray}
\chi_l &=& \frac{1}{V} \sum_x\langle\bar\chi\chi(0)\;\bar\chi\chi(x)
\rangle_c \\
\label{eq:WI}
\chi_t &=& {1\over
V}\sum_x\langle\bar\chi\varepsilon\chi(0)\;\bar\chi
\varepsilon\chi(x)\rangle_c = {1\over
m}\langle\bar\chi\chi\rangle
\end{eqnarray}
They are related to the scalar and pion masses:
\begin{equation}
\label{eq:masses}
\chi_{l,t} = Z_{s,\pi} / M_{s,\pi}^2
\end{equation}
Another interesting quantity is their ratio $R= \chi_l/\chi_t$.
If there is a critical point in the phase diagram where chiral
symmetry is dynamically broken, then:
\begin{eqnarray}
\displaystyle\lim_{m\to0}R=\cases
{0,&$g^2>g_c^2$;\cr {1\over{\delta-{1\over\beta}}},&$g^2<g_c^2$.\cr}
\end{eqnarray}

\begin{figure}[ht]
\vspace{-0.5cm}
\centerline{
\setlength\epsfxsize{200pt}\epsfbox{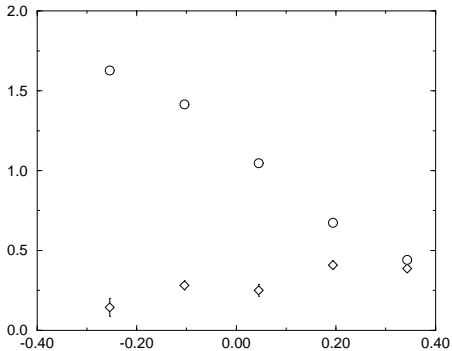}}
\vspace{-1.5cm}
\caption{Transverse (circles) and longitudinal (diamonds)  
susceptibilities for $N_f=4$, $m=0.01$, on a $16^3$ lattice. The data
are plotted against a rescaled reduced coupling $t=g_c^2 (1/g^2 -
1/g_c^2)$.
\label{fig:susc}}
\end{figure}

The results for the susceptibilities are shown in
Fig.~\ref{fig:susc}. Even though the results were produced at
non-vanishing bare mass, i.e. in presence of an explicit symmetry
breaking term, the pattern reflects the existence of a phase
transition. Once again the near degeneracy of $\chi_l$ and $\chi_t$ in
the symmetric phase $(t>0)$ supports the hypothesis
$\delta-1/\beta=1$, used in the fits reported in Tab.~\ref{tab:fit}.

\begin{figure}[htb]
\vspace{-0.5cm}
\centerline{
\setlength\epsfxsize{200pt}\epsfbox{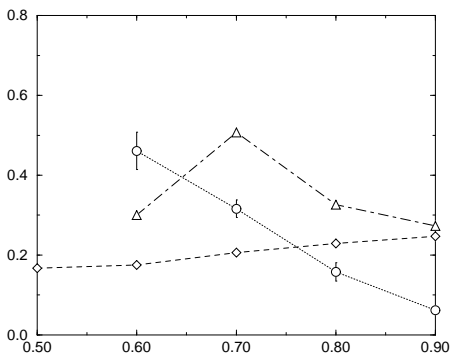}}
\vspace{-1.5cm}
\caption{Fermion (circles), pion (diamonds) and scalar (triangles) 
masses vs. $1/g^2$, for $N_f=4$, $m=0.01$, on a $16^3$
lattice. \label{fig:mass}}
\end{figure}

In the strong coupling regime $(t<0)$, the transverse
susceptibility is larger than the longitudinal, reflecting the fact
that the pion is almost a Goldstone boson in that regime. The same
trend is confirmed by the masses extracted from the aymptotic
behaviour of two--point functions. Preliminary results are reported in
Fig.~\ref{fig:mass}; the errors on the scalar mass is still very
large and has not been included in the plot. A more detailed analysis,
based on larger statistical samples, is currently in progress and we
hope to report soon on the details of the spectrum for $N_f=4$. It is
nonetheless worth to remark that the general pattern is consistent
with the existence of an UV fixed point. In particular, we see no sign
for the unconventional spectrum in the symmetric phase associated to
conformal phase transitions, as suggested in~\cite{mir96}.

\begin{figure}[htb]
\vspace{-0.5cm}
\centerline{
\setlength\epsfxsize{200pt}\epsfbox{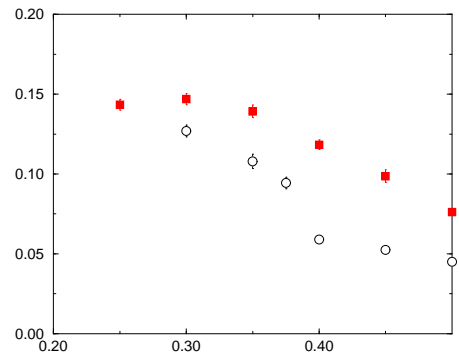}}
\vspace{-1.5cm}
\caption{Chiral condensate vs. $1/g^2$ for $N_f=6$, $m=0.01$ (circles)
on a $16^3$ lattice and $m=0.02$ (squares) on a $12^3$ lattice.
\label{fig:cond}}
\end{figure}

In order to distinguish between different scenarios of chiral symmetry
breaking (corresponding, e.g., to different solutions of the
Schwinger--Dyson equations), several values of $N_f$ need to be
studied. New preliminary results for the chiral condensate are
reported in Fig.~\ref{fig:cond}. The lattice used is larger than in
our earlier works~\cite{ldd_and_sjh} and the behaviour of the
condensate as the bare mass $m$ is decreased suggests the existence of
a symmetry broken phase at strong coupling. We will report more
systematic results in the near future.

{\bf Acknowledgements:} Some of the numerical simulations reported
here are funded under HPCI, EPSRC grant GR/K41663 and PPARC grant
GR/K55745. Last, but not least, it is a pleasure to thank the LOC for
financial support.

\end{document}